# Quantum Machine Learning Using Quantum Illumination With Quantum Enhanced -Interference

Pallab Biswas[1], Tamal Maity[2]
[1,2]The University of Burdwan,WB,India
E-mail: sciencestpallab162@gmail.com[1], tamalmaity89@gmail.com[2]

Quantum Machine Learning (QML) is developed by combining quantum mechanics principles with classical machine learning techniques in a hybrid framework that can give faster,exponential,more efficient power of quantum computing with the data-driven intelligence.Quantum illumination(QI) is the quantum mechanical technique along with analysis of light matter interaction from source to detection end that connects quantum principle to hardware implementation.Superposition and entanglement control are deeply needed for the information–qubit processing in quantum computing. Improvement of measurement and performance are directly linked to detecting weak signal or intensity. This paper motivated that using quantum-enhanced technique how we can analysis previous superposition of qubit state which can clearly analyzed quantum interference - diffraction patterns and its superposition using double slit experiment.Then constructed quantum neural network's back propagation technique such that can give information of qubit position in any previous superposition state.Which is very import for any quantum optimization and search algorithm.

## 1.Introduction

Lord Kelvin said that If you cannot measure it, you cannot improve it. Density or intensity is only measurable things that is directly linked with most valuable quantum properties like superposition and entanglement.Using of quantum illumination ,we deeply study quantum property of light-matter interaction and quantum measurement. QI starts with quantum properties of light from interference,diffraction and its superposition. Quantum computing algorithm [21,22] like Grover's algorithm also discuses interference and superposition principle at the last end.

The concept of interference starts with philosophical observation on anomaly of the tides in the Gulf of Tongkin . In the 17th century, Newton attributed the pattern of the tides of Tongkin to the superposition of component tides that are arriving from different directions[1,2].

In 1801, Thomas Young performed his famous experiment demonstrating that light passing through two very closely spaced slits produces an interference pattern of bright-dark fringes and describe light as a wave[2,3]. After some years,Maxwell and his famous maxwell's equation describe the electromagnetic property of light[6,7]. That is grown light-matter interaction from source to detection end or quantum illumination.

Study of light-matter interaction accomplishment was adaptation by Huygens principle principle of wavelet superposition,Young's conclusion, Fresnel and Fraunhofer mathematical Approximations[5,8].In modern age, Max Planck introduced the concept of energy quantization through his work on black-body radiation, which marked the birth of quantum theory. Arthur compton and Louis de Broglie with helps of X-ray scattering introduced matter wave and wave-particle duality[10,11].

At the last stage ,Werner Heisenberg ,Erwin Schrödinger and Paul A. M. Dirac develop quantum mechanical property of light with photon field quantization,particle annihilation and quantum field theory. Then feynman,Schwingen at all. develop quantum electrodynamics(QED) that properly explain of quantum property of light – interference, scattering, emission and absorption[9,12,13]. That makes heart of quantum illumination.So quantum illumination works on light matter interaction for higher detection and sensitivity which body is made of quantum property of light but heart is made with quantum electrodynamics. Using QI data ,making of a machine learning is the deep task for quantum circuit and quantum computing. This is the brain-body making process which can give extraordinary power to qubit processing of quantum computing.

## 2. Quantum Illumination Port - folio.

Two or more wave with resulting amplitude on about phase relationship contribute the interference of wave from distinct source or path.In scalar wave theory,the redistribution of wave amplitude with every point of waveform makes diffraction. Thus diffraction and interference are the same physical superposition principle works on same wave-function different geometries.

A monochromatic scalar wave from the slit x $\in$ [+ $a/2$,+ $a/2$] ,z$\in$0 is observed at distance than each point on slit with secondary points acts as a uniform illumination[5,10].In double slit experiment, source point with angle θ,path difference between center to observation point will be at position (xsinθ) with phase difference kxsinθ,k=2π/$\lambda$ ,then the complex amplitude will be

$$E(\theta)=\int_{+a/2}^{+a/2} e^{i\,kxsin\theta}dx \ = \frac{2sin(\frac{kasin\theta}{2})}{ksin\theta} \quad ........(1)$$

Now define the parameter $\beta = \frac{kasin\theta}{2} = \frac{\pi asin\theta}{\lambda}$

Then the complex amplitude with constant factor ,

$$E(\theta)=E_{optical}\left(\frac{sin\beta}{\beta}\right) \quad .....(2)$$

where $E_{optical}$ is the optical illumination factor with complex amplitude.

Which is directly depend on intensity of wave and classical single slit diffraction pattern that zero at $\beta=\pm\, mx$ for integer m$\neq$ 0.

At the present of double slit ,each slit is an extended aperture with span of slit-1, x $\in$ [ - $\frac{d}{2}$ - $\frac{a}{2}$, - $\frac{d}{2}$ + $\frac{a}{2}$ ] and slit-2   x $\in$ [$\frac{d}{2}$ - $\frac{a}{2}$, $\frac{d}{2}$ + $\frac{a}{2}$ ] then the complex amplitude.

$$E(\theta) = \int_{-\frac{d}{2}+\frac{a}{2}}^{-\frac{d}{2}-\frac{a}{2}} e^{i\,kxsin\theta}d\theta + \int_{-\frac{d}{2}+\frac{a}{2}}^{-\frac{d}{2}-\frac{a}{2}} e^{i\,kxsin\theta}d\theta \quad .......(3)$$

Sift variable so each integral is same as single slit but with phase from center position.

$$E(\theta) = e^{-ik(\frac{d}{2})sin\theta}\int_{-\frac{d}{2}+\frac{a}{2}}^{-\frac{d}{2}-\frac{a}{2}} e^{i\,kxsin\theta}d\theta +e^{-ik(\frac{d}{2})sin\theta}\int_{-\frac{d}{2}+\frac{a}{2}}^{-\frac{d}{2}-\frac{a}{2}} e^{i\,kxsin\theta}d\theta \quad .....(4)$$

When the single amplitude $E_{optical}$ then the integral composed(for double slit) with

$$E(\theta)=E_{optical}\cdot[e^{-i\alpha}+e^{-i\alpha}] =2.E_{optical}cos\alpha \quad.....(5)$$

With along N identical slit of width a,spacing d ,the total field amplitude is

$$E(\theta)=E_{optical}\sum_{n=0}^{n-1} e^{i\,kndxsin\theta} = E_{optical}\frac{1-e^{iN\phi}}{1-e^{i\phi}}$$

$$= E_{optical}\frac{sin(N\phi/2)}{sin(\phi/2)}, \phi=kdsin\theta \quad ....(6)$$

For the complete equation of quantum illumination with interference and diffraction pattaren .

$$E(\theta)=E_{optical}[\frac{sin\beta}{\beta}][\frac{sinN\alpha}{\alpha}] \quad .......(7)$$

In the complete wave picture, every point on a wavefront acts as a secondary source of new waves, and the final observed pattern at a screen or detector emerges from the coherent superposition of all these secondary wavelets and all forms of wave motion propagate, spread, and form structured intensity distributions in space.

## 3. Bloch Sphere and Optical Qubit Combination

A single qubit exists in a superposition of $|0\rangle$ and $|1\rangle$ and The Bloch sphere is a geometrical representation of the state space of a two-level quantum system (a qubit).In 1946, Felix Bloch introduced Bloch equations which given macroscopic magnetization as a vector in three-dimensional space, precessing under applied fields and relaxing due to interactions.Feynman, Vernon, and Hellwarth in 1957 showed that the Schrödinger equation for a two-level system can be mapped to a vector equation of the form dr/dt = ω × r, where r is a three-dimensional real vector whose components correspond to real combinations of the probability amplitudes of the two quantum states.The two angles (θ, φ) where θ = polar angle,(θ∈0,π),φ = azimuthal angle,(φ∈ 0,2π) on the sphere correspond to the two independent real parameters (up to a global phase) of a qubit's pure state.The Bloch sphere maps every pure qubit state to a point on the surface of a unit sphere.Single-qubit quantum gates (unitary operations) correspond to rotations of the Bloch vector around the sphere[14,15].For mixed states, the length of the Bloch vector (less than 1) gives a visual measure of purity: pure states are on the surface, fully mixed states are at the center. Bloch-sphere description equation of a qubit is.

$$|\psi\rangle = cos\,\theta'/2\,|0\rangle + e^{i\varphi'}\,sin\,\theta'/2\,|1\rangle \quad ....(8)$$

When two photon or qubit ($|\psi_1\rangle$, $|\psi_2\rangle$) pass through from both individual slit (slit-1 & slit -2)on a single unit of time.Then their combination is a subtle quantum process where two individual photon behavior depends on how their wavefunctions overlap. Two photon or qubit combination gives amplitude and relative phase information that is responsible for quantum illumination. This quantum illumination farther is constructed intensity layer that's final products are fringes,interference, diffraction and also scattering. then their wavefunctions ($|\psi_1\rangle$, $|\psi_2\rangle$) with bloch-sphere equation will be

$$|\psi_1\rangle = cos\,\frac{\theta'_1}{2}|0\rangle + e^{i\phi'_1}\,sin\frac{\theta'_1}{2}\,|1\rangle \quad ...(9)$$

$$|\psi_2\rangle = cos\,\frac{\theta'_2}{2}|0\rangle + e^{i\phi'_2}\,sin\frac{\theta'_2}{2}\,|1\rangle \quad .....(10)$$

This two qubit combination on computation basis happens through tensor products and system to hold correlations on four possible states $|00\rangle$, $|01\rangle$, $|10\rangle$, $|11\rangle$.

Each basis state captures a full pairing between the first qubit's outcome and the second qubit's outcome. And because quantum mechanics allows superpositions of all these basis states, the two-qubit system can occupy all four combinations simultaneously, each with its own complex-valued probability amplitude.

$$|\psi\rangle = |\psi_1\rangle \otimes |\psi_2\rangle \quad ......(11)$$

Then $|\psi\rangle = cos\,\frac{\theta'_1}{2}cos\,\frac{\theta'_2}{2}|00\rangle + cos\,\frac{\theta'_1}{2}e^{i\phi'_2}sin\frac{\theta'_2}{2}$
$|01\rangle + e^{i\phi'_1}\,sin\frac{\theta'_1}{2}cos\,\frac{\theta'_2}{2}\,|10\rangle +$
$e^{i(\phi'_1+\phi'_2)}\,sin\frac{\theta'_1}{2}sin\frac{\theta'_2}{2}|11\rangle \quad .....(12)$

In quantum mechanics ,intensity is proportional to modulus square of its amplitude. So the table of two qubit combination.

| Base state | Amplitude of qubit | Intensity |
|---|---|---|
| $|00\rangle$ | $cos\,\frac{\theta'_1}{2}\,cos\,\frac{\theta'_2}{2}$ | $I_{00}= cos^2\,\frac{\theta'_1}{2}\,cos^2\,\frac{\theta'_2}{2}$ |
| $|01\rangle$ | $cos\,\frac{\theta'_1}{2}e^{i\phi'_2}sin\frac{\theta'_2}{2}$ | $I_{01}= cos^2\frac{\theta'_1}{2}sin^2\frac{\theta'_2}{2}$ |
| $|10\rangle$ | $e^{i\phi'_1}\,sin\frac{\theta'_1}{2}cos\,\frac{\theta'_2}{2}$ | $I_{10}= sin^2\frac{\theta'_1}{2}cos^2\,\frac{\theta'_2}{2}$ |
| $|11\rangle$ | $e^{i(\phi'_1+\phi'_2)}\,sin\frac{\theta'_1}{2}sin\frac{\theta'_2}{2}$ | $I_{11}= sin^2\frac{\theta'_1}{2}sin^2\frac{\theta'_2}{2}$ |

But for the interference pattern relative phase between two superposed component plays an important role for making fringes.

$$|\psi\rangle = I_{\psi_1}e^{\phi'_1} + I_{\psi_2}e^{\phi'_2} \quad \ldots\ldots(13)$$

The probability distribution (or probability amplitude squared) associated with the combined quantum state.

Total intensity for qubit combination.

$$I = |\psi|^2 = I_{\psi_1}e^{\phi'_1} + I_{\psi_2}e^{\phi'_2})(I_{\psi_1}e^{-\phi'_1} + I_{\psi_2}e^{-\phi'_2})$$

The total intensity always adds up to 1, because probabilities must sum to 1. But the way this intensity is distributed across the four computational basis states depends entirely on how the qubits combine through tensor products.

So, $I_{00} + I_{01} + I_{10} + I_{11} = 1$ ...(14)

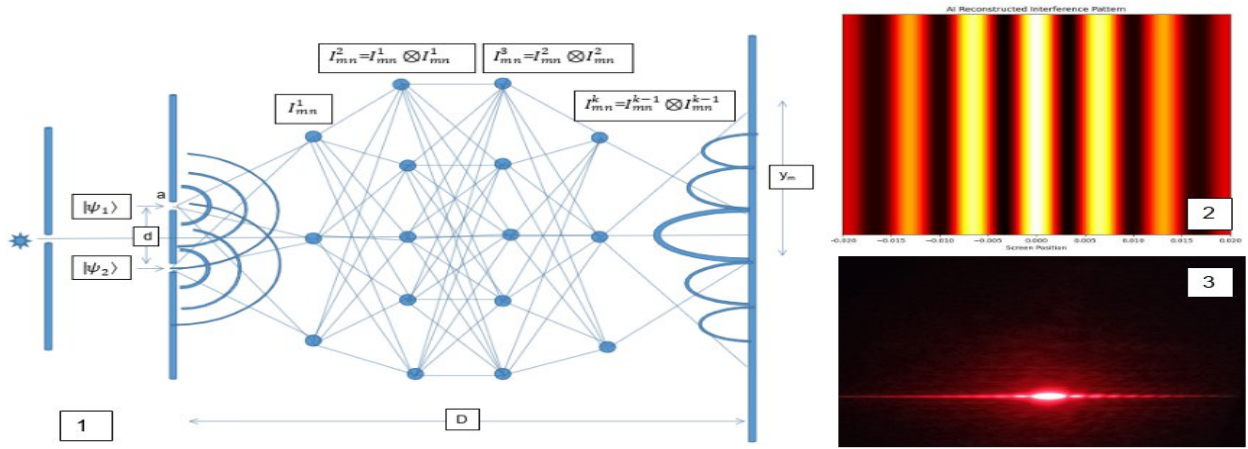

**Fig 1: (1) Double slit experiment and quantum machine learning model with qubit combination. (2) Simulation based quantum interference. (3) Quantum interference laboratory images which data is used for Training of QNN.**

$$= I_{\psi_1}^2 + I_{\psi_2}^2 + 2I_{\psi_1}I_{\psi_2}\cos\phi'$$

$$= (I_{\psi_1}^2 + I_{\psi_2}^2)[1 + \frac{2I_{\psi_1}I_{\psi_2}}{I_{\psi_1}^2 + I_{\psi_2}^2}\cos\phi'] \quad \ldots(15)$$

$= I_{mn}[1 + v\cos\phi']$ where v is the coherence property of light, and $\phi' = \phi'_2 - \phi'_1$, and

$I_{mn} = (I_{\psi_1}^2 + I_{\psi_2}^2)$ is the intensity amplitude of qubits combination.

If the wave is perfectly coherence, then

For v=1

$$I = I_{mn}\cos^2(\frac{\phi'}{2})$$

So, $E_{optical} = I_{mn}\cos^2(\frac{\phi'}{2})$ ......(17) and intensity due to interference and diffraction is that

$[\frac{\sin\beta}{\beta}]^2[\frac{\sin N\alpha}{\alpha}]^2$.

Wave-function and it's final equation

$$E(\theta,\alpha,\beta,\phi) = 2.I_{mn}\cos^2(\frac{\phi'}{2})[\frac{\sin\beta}{\beta}]^2[\frac{\sin N\alpha}{\alpha}]^2$$

$$\ldots\ldots(18)$$

Where, $I_{mn} = \begin{bmatrix} I_{00} \\ I_{01} \\ I_{10} \\ I_{11} \end{bmatrix}$, $\alpha = \frac{\pi d\sin\theta}{\lambda}$, $\beta = \frac{\pi a\sin\theta}{\lambda}$, N=2 (no of slit) and another parameter slit separation d, slit width a, wavelength $\lambda$, and diffraction angle $\theta$.

## 4. Energy distribution optimization for Quantum Machine Learning

Quantum Machine Learning (QML) is an emerging field that merges quantum computing with machine-learning techniques to process information in ways impossible for classical systems. By using quantum principles such as superposition, entanglement, and interference, QML algorithms can explore large data spaces more efficiently and provide faster pattern recognition or optimization for complex problems. Within QML, Quantum Neural Networks (QNNs) act as quantum analogs of classical neural networks, where quantum circuits play the role of layers and tunable quantum gates act as learnable parameters.

advantages in speed, accuracy, and modeling capacity, opening new pathways for intelligent computing and next-generation AI powered by quantum technologies.

Quantum Neural Networks (QNNs) can be used to analyze and optimize the interference and

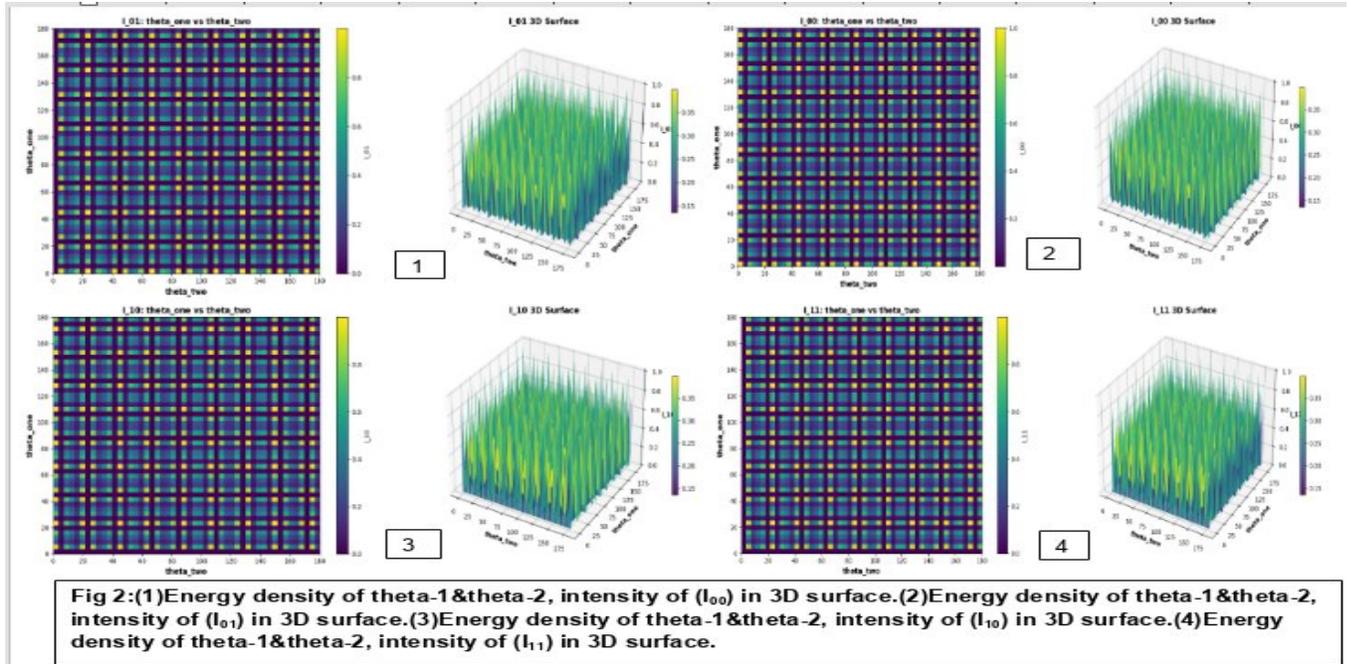

Fig 2: (1) Energy density of theta-1 & theta-2, intensity of ($I_{00}$) in 3D surface. (2) Energy density of theta-1 & theta-2, intensity of ($I_{01}$) in 3D surface. (3) Energy density of theta-1 & theta-2, intensity of ($I_{10}$) in 3D surface. (4) Energy density of theta-1 & theta-2, intensity of ($I_{11}$) in 3D surface.

These networks can encode data into quantum states and use quantum transformations to detect patterns that classical models may find too large or too nonlinear to handle.

diffraction patterns produced in Young's double-slit experiment, where light or matter waves create fringes due to superposition. In this approach, the interference data such as fringe spacing, intensity distribution, phase shifts, and QNNs are especially powerful for tasks involving high-dimensional feature spaces, quantum physics simulations, and noise-resilient learning. Together, QML and QNNs promise significant slit geometry—is encoded into quantum states and processed through a QNN, which uses entanglement and quantum gates to learn hidden patterns in wave behavior. The QNN can adjust parameters like slit width, separation, wavelength, coherence and detection-plane distance to automatically optimize the resulting pattern for higher contrast, improved visibility, or desired symmetry. By exploiting quantum parallelism, the network explores many possible configurations at once, giving faster convergence than classical models. This combination of quantum learning with a fundamental interference experiment enables advanced control of optical qubit.

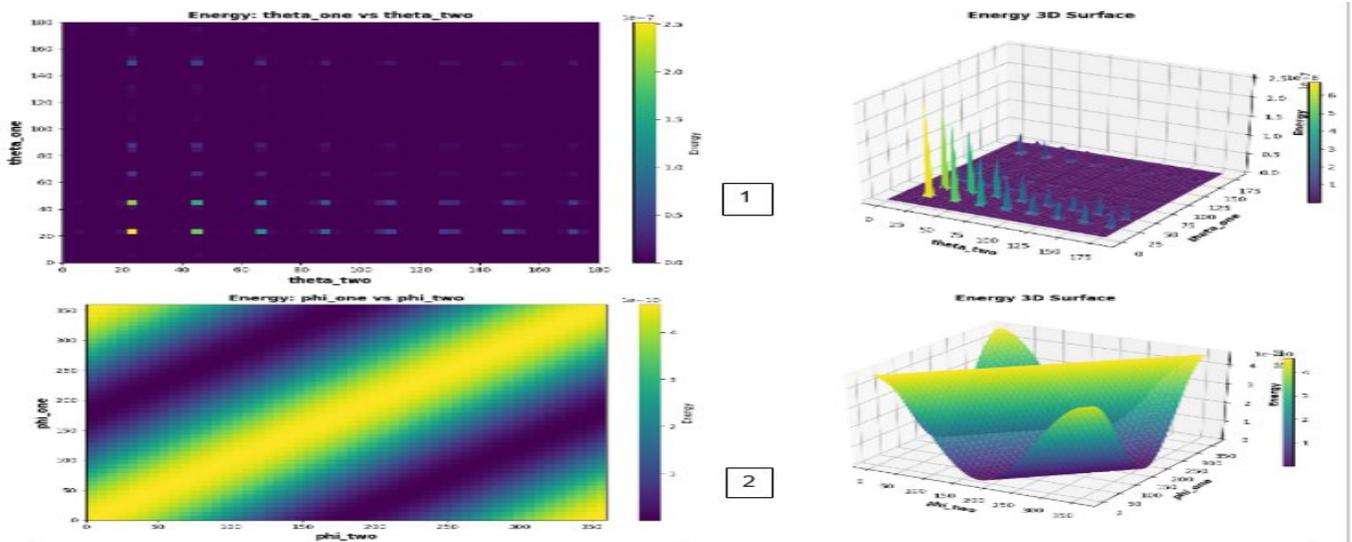

Fig 3: (1) Total Energy density of theta-1 & theta-2 in 3D surface. (2) Total Energy density of phi-1 & phi-2 in 3D surface.

QNN is fundamentally connected through the concept of energy-based optimization and computational problem is encoded into a Hamiltonian, and the goal is to drive the quantum system toward states that minimize the energy expectation value. QNN rely on the redistribution of quantum probability amplitudes across energy eigenstates, the exploitation of superposition and entanglement, and the tendency of quantum systems to favor low-energy configurations.

In a two-qubit version of the double-slit experiment, the system is extended from a single quantum particle to two correlated quantum degrees of freedom. this configuration allows joint interference patterns that depend on correlations between the qubits( $|\psi_1\rangle, |\psi_2\rangle$) and four different parameters $\{\theta_1, \phi_1, \theta_2, \phi_2\}$ in a four-dimensional joint Hilbert space, allowing superposition and entanglement between the two paths. the Bloch-sphere angles($\theta_1, \theta_2$) control energy distribution, while entanglement and phase($\phi_1, \phi_2$) control interference and information.

QNN prepares and processes information by evolving an quantum state through a parameterized unitary transformation U(θ) that is is analogous to weights in a classical neural network.

$$|\psi(\theta_1,\phi_1,\theta_2,\phi_2)\rangle = U(\theta_1,\phi_1,\theta_2,\phi_2)|0\rangle + U(\theta_1,\phi_1,\theta_2,\phi_2)|1\rangle \quad ....(19)$$

Physical observable is represented by a Hermitian operator, and energy is represented by the Hamiltonian(H).Fig-2.

Quantum models evaluate performance using the energy expectation value, which represents the average energy of a quantum state with respect to a problem-specific Hamiltonian.(Fig 3)

Energy Expectation Value

$$E = \langle\psi(\theta_1,\phi_1,\theta_2,\phi_2)|H|\psi(\theta_1,\phi_1,\theta_2,\phi_2)\rangle .....(20)$$

Loss function is defined as

$$\zeta(\theta_1,\phi_1,\theta_2,\phi_2) = \langle\psi(\theta_1,\phi_1,\theta_2,\phi_2)|H|\psi(\theta_1,\phi_1,\theta_2,\phi_2)\rangle$$

$$..............(21)$$

Training of QNN depend on

$$(\theta_1,\phi_1,\theta_2,\phi_2)^* = \arg\min \langle\psi(\theta_1,\phi_1,\theta_2,\phi_2)|H|\psi(\theta_1,\phi_1,\theta_2,\phi_2)\rangle \quad ...(22)$$

Using the final equation (18), is calculated the neural pulse and wave function E(θ,α,β,φ). The first layer of intensity matrix for neural network

$$I^1_{mn} = \begin{bmatrix} I^1_{00} & I^1_{01} & I^1_{10} & I^1_{11} \end{bmatrix} \begin{bmatrix} I^2_{00} \\ I^2_{01} \\ I^2_{10} \\ I^2_{11} \end{bmatrix}$$

$$= \begin{bmatrix} I^1_{00}I^2_{00} & I^1_{00}I^2_{01} & I^1_{00}I^2_{10} & I^1_{00}I^2_{11} \\ I^1_{01}I^2_{00} & I^1_{01}I^2_{01} & I^1_{01}I^2_{10} & I^1_{01}I^2_{11} \\ I^1_{10}I^2_{00} & I^1_{10}I^2_{01} & I^1_{10}I^2_{10} & I^1_{10}I^2_{11} \\ I^1_{11}I^2_{00} & I^1_{11}I^2_{01} & I^1_{11}I^2_{10} & I^1_{11}I^2_{11} \end{bmatrix} ...(23)$$

The second layer of intensity matrix for neural network

$$I^2_{mn} = I^1_{mn} \otimes I^1_{mn} \quad ........(24)$$

For the third layer of intensity matrix

$$I^3_{mn} = I^2_{mn} \otimes I^2_{mn} \quad .......(25)$$

For the n layer of of intensity matrix for neural network will be

$$I^k_{mn} = I^{k-1}_{mn} \otimes I^{k-1}_{mn} \quad .......(26)$$

Back propagation and superposition identification work together to optimize the resulting interference and diffraction patterns. The quantum superposition created when light or particles pass through both slits is encoded directly into quantum states that serve as inputs to the QNN. The network analyzes how variations in slit width, spacing, coherence, and wavelength affect fringe visibility, spacing, and symmetry.

## 5. Quantum Machine Learning Framework for Optical Control

### 5.1 Problem Formulation

Classic interference experiments rely on fixed geometry and static optical sources. In our proposed framework, we treat the optical setup as a programmable quantum circuit. The objective of the Quantum Machine Learning (QML) model is to solve the inverse diffraction problem: determining the precise qubit state configurations $(|\psi_1\rangle, |\psi_2\rangle)$ required to produce a specific target intensity distribution $I_{target}(\Theta)$ at the detection plane.

We define the control parameters as the Bloch sphere angles for the two illuminating qubits:

$$\Phi = \{\theta_1, \phi_1, \theta_2, \phi_2\}$$

These parameters directly modulate the amplitude balance $I_{mn}$ and the relative phase difference $(\triangle \phi')$ in the system's governing wave equation (Eq. 18).

### 5.2 Quantum Neural Network (QNN) Architecture

We construct a **hybrid quantum-classical neural network** where the physical diffraction process acts as a fixed, non-linear transformation layer. The architecture consists of three stages:

1. **Encoding Layer (Variational Circuit):** The system initializes two qubit states using rotation gates $R_y(\theta)$ and $R_z(\phi)$. These gates serve as the trainable weights of the network. The quantum state of the system is represented as the tensor product of the individual slit wavefunctions.

2. **Diffraction Layer (Physical Model):** Unlike standard artificial neural networks that use sigmoid or *ReLU activation functions*, our "activation function" is the physical interference equation derived in Section 3:

$$E(\Theta, \Phi) = 2\, I_{mn}(\Phi)\, \cos^2\left(\frac{\phi'(\Phi)}{2} + \alpha\right) \mathrm{sinc}^2(\beta) \quad \ldots (27)$$

This layer is fully differentiable, allowing us to compute gradients of the output intensity with respect to the input Bloch angles.

3. **Cost Function and Optimization:** To achieve active beam steering, we define a utility-based loss function $\mathcal{L}$. For a desired target angle $\Theta_{target}$ the loss is defined to maximize intensity at that specific location:

$$\mathcal{L}(\phi) = -E(\Theta_{target}, \phi) \quad \ldots(28)$$

The network minimizes this loss using the Adam optimizer, a gradient-descent algorithm that iteratively updates $\phi$ to align the constructive interference fringes with $\Theta_{target}$.

### 5.3 Learning Algorithm

The training process follows a "Physics-Informed" learning loop. In each epoch k:

1. **Forward Pass:** The QNN computes the resulting interference pattern $E_k(\Theta)$ based on current parameters $\Phi_k$.

2. **Gradient Computation:** The system calculates the partial derivatives of the intensity field with respect to the phase angles:

$$\nabla \mathcal{L} = \left[\frac{\partial \mathcal{L}}{\partial \theta_1}, \frac{\partial \mathcal{L}}{\partial \phi_1}, \frac{\partial \mathcal{L}}{\partial \theta_2}, \frac{\partial \mathcal{L}}{\partial \phi_2}\right] \quad ..(29)$$

3. **Backpropagation:** The parameters are updated to move the system state closer to the optimal interference condition:

$$\Phi_{k+1} = \Phi_k - \eta\, \nabla \mathcal{L} \quad ..(30)$$

Where, η is the learning rate

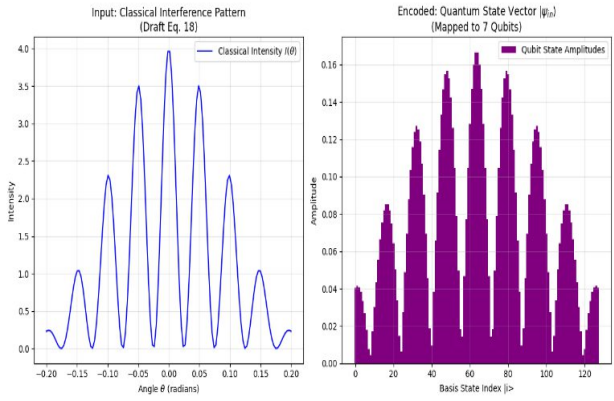

**Fig4: Classical interference pattern v/s. Encoded Quantum State**

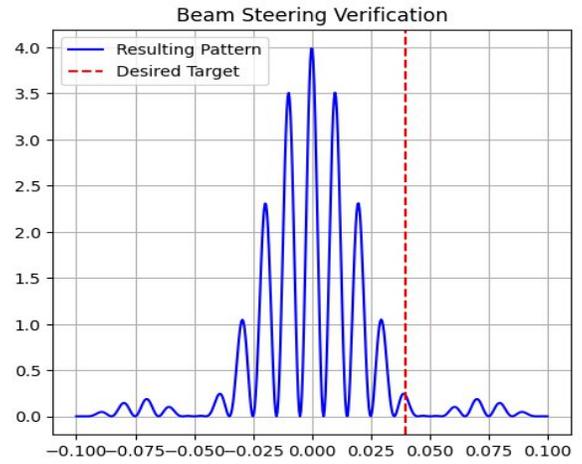

**Fig. 5(a): Our QNN model training and it's simulation results**

The Quantum Machine Learning model was implemented in Python using the **PyTorch** deep learning framework **[11]**, which allows for automatic differentiation of the physical wave equations. The numerical simulation of the diffraction patterns utilized **NumPy** for high-performance array computing **[12]**.

The optimization of the Bloch sphere parameters ($\theta, \phi$) was performed using the **Adam** (Adaptive Moment Estimation) optimizer **[13]**. This gradient-based optimization algorithm was selected for its efficiency in handling non-convex objective functions and its adaptive learning rate capabilities, which are crucial for navigating the complex energy landscape of optical interference patterns.

**5.4 Simulation Results**

The QML model was trained to focus maximum intensity at an arbitrary off-axis angle of $\theta = 0.04$ rad. As shown in Fig. 5, the optimization converged within 100 epochs.

The network successfully identified a phase configuration where the relative phase difference $\Delta\phi \approx \pi$ rad. While the global diffraction envelope (governed by slit width $a$) remains static, the QNN shifted the internal interference fringes such that a principal maximum aligned perfectly with the target coordinate.

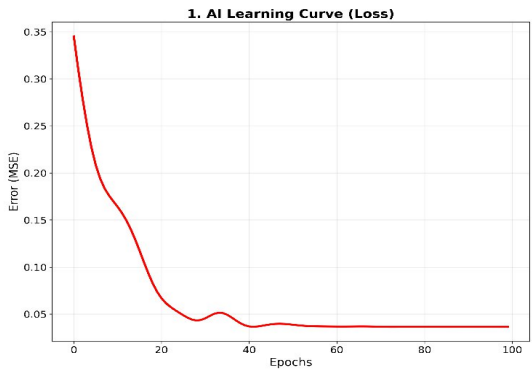

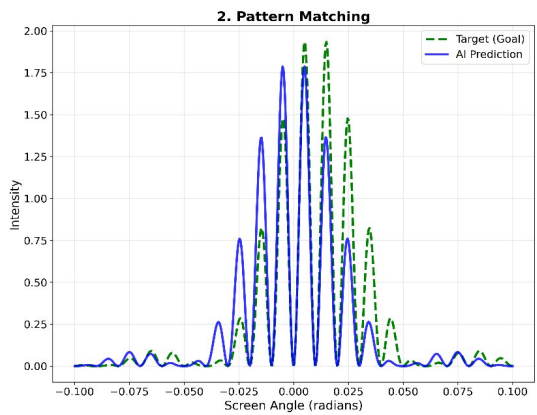

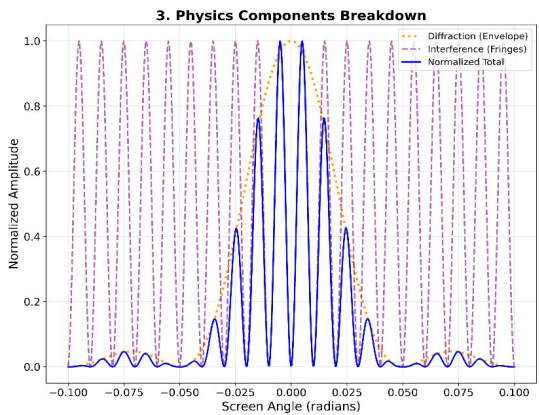

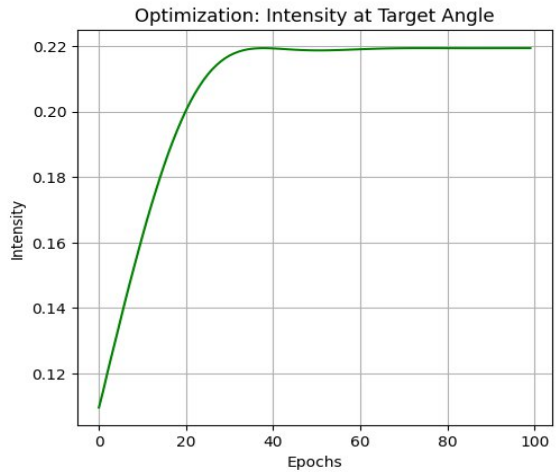

**Fig. 5(b): Active Beam Steering with our QNN model**

This demonstrates that the system can perform **"Fringe Steering"** —autonomously tuning the quantum states of the source to maximize signal detection at specific spatial coordinates.

The optimization demonstrates **'Fringe Steering'** rather than **'Envelope Steering'**. As seen in Figure 5(b), the global diffraction envelope (governed by the fixed slit width $a$) restricts the maximum potential intensity at large angles. However, the QNN successfully adjusted the relative phase $\Delta\phi$ to align a constructive interference maximum exactly with the target angle of 0.04 rad. This confirms the system's ability to maximize signal detection at off-axis locations within the constraints of the physical diffraction limit.

When superposition is preserved and detected, it confirms that the system is truly behaving according to quantum mechanics and that coherence has not been destroyed by noise or decoherence. This ability enables quantum parallelism, where multiple solution paths are processed at once, and it also makes quantum

## Training History (Every Epoch)

This table tracks the evolution of Bloch angles, Intensities, and Loss over the entire training process.

| Epoch | Loss Function | Theta-1 ($\theta_1$) | Phi-1 ($\phi_1$) | Theta-2 ($\theta_2$) | Phi-2 ($\phi_2$) | Possible states Amp. $|00\rangle$ | Possible states Amp. $|01\rangle$ | Possible states Amp. $|10\rangle$ | Possible states Amp. $|11\rangle$ |
|---|---|---|---|---|---|---|---|---|---|
| 1 | 0.34517795 | 1.6708 | 0.1000 | 1.6708 | -0.1000 | 0.2026 | 0.2475 | 0.2475 | 0.3024 |
| 2 | 0.31043187 | 1.7700 | 0.1746 | 1.7700 | -0.1746 | 0.1608 | 0.2402 | 0.2402 | 0.3587 |
| 3 | 0.27917820 | 1.8674 | 0.2596 | 1.8674 | -0.2596 | 0.1252 | 0.2286 | 0.2286 | 0.4175 |
| 4 | 0.25123212 | 1.9616 | 0.3496 | 1.9616 | -0.3496 | 0.0958 | 0.2137 | 0.2137 | 0.4768 |
| 5 | 0.22752340 | 2.0506 | 0.4430 | 2.0506 | -0.4430 | 0.0725 | 0.1967 | 0.1967 | 0.5341 |
| 6 | 0.20855235 | 2.1319 | 0.5389 | 2.1319 | -0.5389 | 0.0547 | 0.1792 | 0.1792 | 0.5868 |
| 7 | 0.19422016 | 2.2028 | 0.6367 | 2.2028 | -0.6367 | 0.0419 | 0.1628 | 0.1628 | 0.6326 |
| 8 | 0.18381453 | 2.2608 | 0.7357 | 2.2608 | -0.7357 | 0.0330 | 0.1487 | 0.1487 | 0.6696 |
| 9 | 0.17617613 | 2.3042 | 0.8353 | 2.3042 | -0.8353 | 0.0273 | 0.1380 | 0.1380 | 0.6967 |
| 10 | 0.16998497 | 2.3321 | 0.9349 | 2.3321 | -0.9349 | 0.0240 | 0.1310 | 0.1310 | 0.7139 |
| 20 | 0.07377753 | 1.9627 | 1.7646 | 1.9627 | -1.7646 | 0.0955 | 0.2135 | 0.2135 | 0.4775 |
| 30 | 0.04383838 | 1.4569 | 1.5456 | 1.4569 | -1.5456 | 0.3101 | 0.2468 | 0.2468 | 0.1964 |
| 40 | 0.03816766 | 1.6035 | 1.5620 | 1.6035 | -1.5620 | 0.2339 | 0.2497 | 0.2497 | 0.2666 |
| 50 | 0.03923010 | 1.7370 | 1.6048 | 1.7370 | -1.6048 | 0.1741 | 0.2432 | 0.2432 | 0.3396 |
| 60 | 0.03688146 | 1.6223 | 1.5421 | 1.6223 | -1.5421 | 0.2249 | 0.2493 | 0.2493 | 0.2764 |
| 70 | 0.03672274 | 1.6309 | 1.5886 | 1.6309 | -1.5886 | 0.2209 | 0.2491 | 0.2491 | 0.2809 |
| 80 | 0.03660397 | 1.6661 | 1.5593 | 1.6661 | -1.5593 | 0.2047 | 0.2477 | 0.2477 | 0.2998 |
| 90 | 0.03647941 | 1.6384 | 1.5768 | 1.6384 | -1.5768 | 0.2174 | 0.2489 | 0.2489 | 0.2849 |
| 100 | 0.03645585 | 1.6470 | 1.5674 | 1.6470 | -1.5674 | 0.2134 | 0.2485 | 0.2485 | 0.2895 |

interference possible, which is the mechanism through which correct answers are amplified and incorrect ones are suppressed in quantum algorithms.

Our final final optimized parameters for this simulation of the two illuminating qubits are :

$$\Phi = \{\theta_1 = 1.6470, \phi_1 = 1.5674, \theta_2 = 1.6470,$$

$$\phi_2 = -1.5674\}$$

Final Two-Qubit State with phase

$|\psi\rangle = $ 0.4624$|00\rangle$+(0.0017−i0.4986)$|01\rangle$
 +(0.0017+i0.4986)$|10\rangle$+0.5376$|11\rangle$

And final Two-Qubit Amplitude ( intensity) state.

$|\psi\rangle = $ 0.2134$|00\rangle$+0.2485$|01\rangle$
 +0.2485$|10\rangle$+0.2895$|11\rangle$

It is the unique superposition value of quantum enhanced interference and unique solution of quantum search problem.

## Summary

Quantum Machine Learning using Quantum Illumination in the context of the double-slit experiment focuses on learning and enhancing interference patterns created by optical qubit. In this approach, quantum illumination provides correlated optical qubit that pass through the slits, allowing the machine-learning model to analyze tiny changes in fringe visibility.

This paper presented a novel Quantum Machine Learning (QML) framework for controlling optical interference patterns through "Quantum Illumination" of a double-slit system. By modeling the illuminating source as a pair of tunable qubits, we derived a transfer function linking Bloch sphere parameters $(\theta, \phi)$ to the far-field diffraction intensity.

We implemented a Quantum Neural Network (QNN) to solve the inverse diffraction problem. Simulation results demonstrated that the network could successfully learn the complex phase relationships required to manipulate the interference fringes. Specifically, the QNN identified that a phase difference of $\Delta\phi \approx \pi$ allows for active "Fringe Steering," shifting the intensity maximum to a desired off-axis target angle without mechanical movement.

This work bridges the gap between fundamental wave mechanics and modern learning algorithms, suggesting new applications in programmable optics, secure quantum communication, and adaptive sensing where precise control of light-matter interaction is required.